\documentclass[conference]{IEEEtran}
\usepackage[utf8]{inputenc}
 
\usepackage{tikz}
\usepackage{amsmath}
\usepackage{url}
\usepackage{breakurl}

\begin{document}
\title{The Smart$^2$ Speaker Blocker: An Open-Source Privacy Filter for Connected Home Speakers}
\makeatletter
\newcommand{\linebreakand}{%
  \end{@IEEEauthorhalign}
   \hfill\mbox{}\par
   \mbox{}\hfill\begin{@IEEEauthorhalign}
}
\makeatother

 \author{
 \IEEEauthorblockN{Ilesanmi Olade}
 \IEEEauthorblockA{Liverpool University\\
 Ilesanmi.olade@liverpool.ac.uk}
 \and
 \IEEEauthorblockN{Christopher Champion}
 \IEEEauthorblockA{Xi'an Jiaotong-Liverpool University\\
 cchampion@student.xjtlu.edu.cn}
\and
 \IEEEauthorblockN{Haining Liang}
 \IEEEauthorblockA{Xi'an Jiaotong-Liverpool University\\
 haining.liang@xjtlu.edu.cn}
\linebreakand
 \IEEEauthorblockN{Charles Fleming}
 \IEEEauthorblockA{University of Mississippi\\
 fleming@olemiss.edu}
 }





 


\maketitle
\begin{abstract}
The popularity and projected growth of in-home smart speaker assistants, such as Amazon's Echo, has raised privacy concerns among consumers and privacy advocates. As part of the normal operation of the smart speaker, audio data from the user's home is constantly streamed to remote servers. Notable questions regarding the collection and storage of this data by for-profit organizations include: what data is being collected and how is it being used, who has or can obtain access to such data, and how can user privacy be maintained while providing useful services. In addition to concerns regarding what the speaker manufacturer will do with your data, there are also more fundamental concerns about the security of these devices, third-party plugins, and the servers where they store recorded data. 

To address these privacy and security concerns, we introduce a first-of-its-kind intermediary device to provide an additional layer of security, which we call the \textit{smart, smart speaker blocker} or Smart\textsuperscript{2} for short. By intelligently filtering sensitive conversations, and completely blocking this information from reaching a smart speaker's microphone(s), the Smart$^2$ Speaker Blocker is an open-source, \textbf{offline} smart device that provides users with decisive control over what data leaves their living room. If desired, it can completely hide any identifying characteristics by only allowing the smart speaker to hear a synthesized text-to-speech voice
\end{abstract}

\section{Introduction}
\begin{figure*}[ht]

\centering
\includegraphics[width=0.8\textwidth]{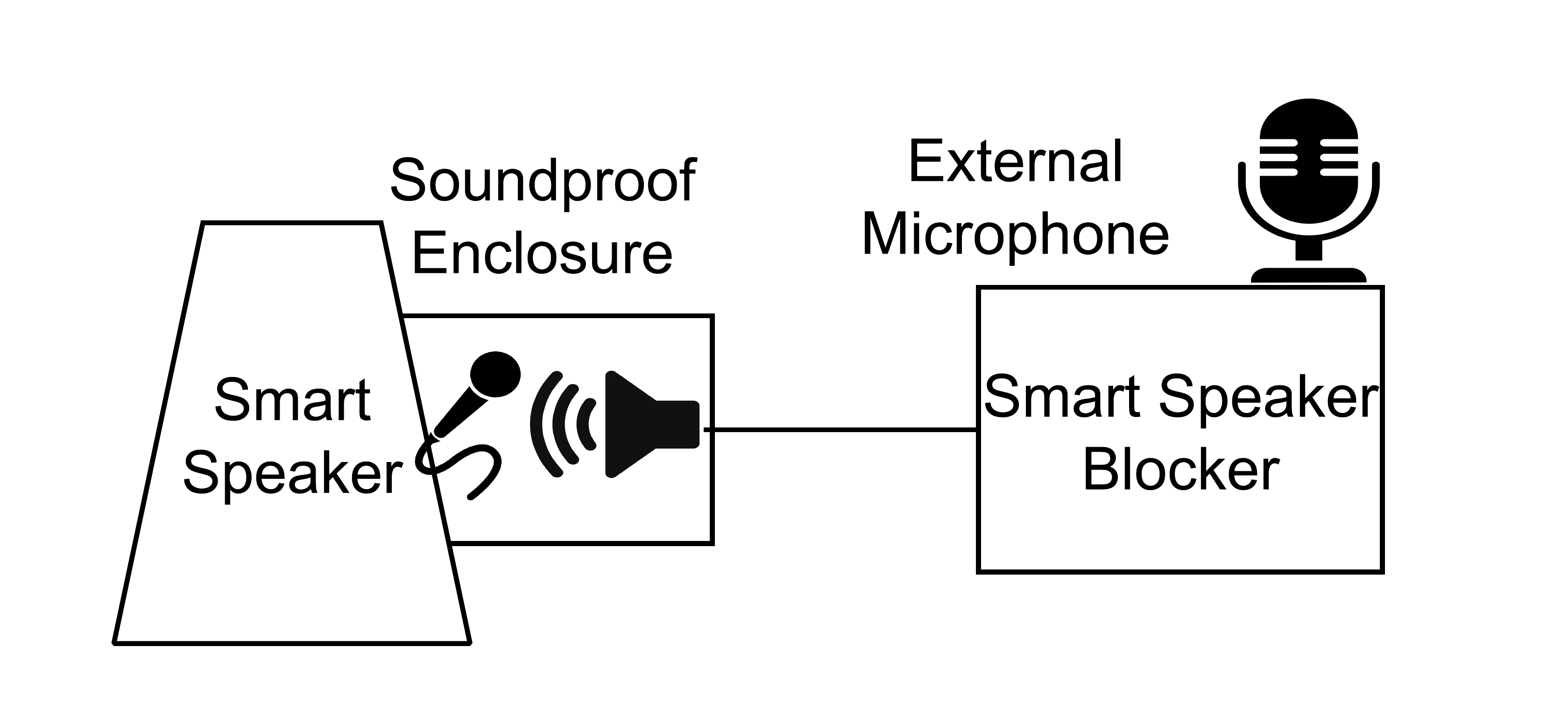}
\caption{The microphone of the smart speaker is insulated from external sounds, and can only hear sounds forwarded by the smart speaker blocker.}
\label{systemDiagram}
\end{figure*}

Market research has shown rampant growth in the adoption of smart speakers with built-in digital assistants, such as Amazon's Alexa or Google's Home line of products, with more than 43 million adults in the United States owning a smart speaker as of spring 2018~\cite{NPR_Edison}. A September 2018 survey estimated 57.8 million Americans own at least one smart speaker, representing 22\% overall domestic growth for the sector~\cite{voicebot_ai1}. Market newcomers, such as Facebook's Alexa-enabled Portal devices, and the continued expansion of product portfolios with a lower cost of entry have forecast a worldwide install base of 100 million by the end of 2018, a number projected to more than double to 225 million by 2020~\cite{canalys}. 

Along with this growth in popularity has come a growing concern about the privacy implications of these devices. Their always on nature, coupled with the fact that they stream the recorded audio to manufacturer-owned servers for voice recognition means that they potentially give these companies access to almost everything that goes on in the owners' homes.  What this data can be used for, who has access to it, and how long it is stored is often vague, and buried in a lengthy user agreement that the user consents to when they first use the device, and is subject to revision at any time.  In addition, many of these devices support extended functionality via plugin modules written by third party developers, who have their own data usage and retention agreements.  The end result is that who has access to recorded data and how it is used is often opaque to the the end user.

Concerns regarding how data is used by legitimate parties aside, there are also concerns regarding the security of these devices. Smart speakers are typically low cost devices, and are produced by dozens of manufacturers. Based on past experience, it is likely only a matter of time before one or more of these devices are compromised, with potentially serious ramifications.  For example, one recorded phone conversation with the user's bank would provide an attacker with all the information necessary to access the user's bank funds.  Alternately, a data breach of a large manufacturer's servers could leak recorded conversations for millions of users, allowing the attacker to perform anything from blackmail to identity theft.

In this paper we describe the first proposed solution to this problem: the \textit{smart, smart speaker blocker}, or Smart\textsuperscript{2} for short. The Smart\textsuperscript{2} works by enclosing the smart speaker's built-in microphone in a soundproof enclosure, along with an external speaker connected to the Smart\textsuperscript{2} device (see Figure \ref{systemDiagram}).  The Smart\textsuperscript{2} has its own microphone and records audio, which it filters using speech-to-text, based on user configured parameters.  Audio that does not match the filter rules is played back for the smart speaker via the external speaker.  Several different filtering modes are implemented, to address a variety of security concerns, including the ability to completely hide any identifying characteristics by only allowing the smart speaker to hear a synthesized text-to-speech voice. The Smart\textsuperscript{2} is air-gapped (not connected to an external network), making it difficult for external attackers to compromise, with all processing done on the local device.  All software, including the speech-to-text library, is open-source and under the end user's control, providing complete transparency.

 Because this is the first system to protect smart speakers from accidental data leakage, there are no competing systems to benchmark against. Instead, our evaluation of the Smart\textsuperscript{2} consists of two parts: system performance impact and a security evaluation.  For the system performance impact, we measure the impact of the Smart\textsuperscript{2} on the smart speaker performance, particularly the introduction of latency by the speech-to-text engine, which we demonstrate is minimal.  To evaluate the enhanced security offered by the Smart\textsuperscript{2}, we perform a two week study of the device's performance in a real life environment and show that it substantially reduces the amount of sensitive audio data leaked, while minimally impacting the usability of the device. 

\section{Background and Related Work}

The popularity of home voice assistants has caught the attention of many digital rights organizations and privacy advocates such as Digitalcourage, the organizing body of the \textit{Big Brother Awards} in Germany~\cite{BigbroAwards1}. The group awarded the 2018 ``Consumer Protection'' category winner to Amazon Alexa for Amazon's practice of storing recorded user interactions on remote servers~\cite{BigbroAwards2}. They were also especially critical of Amazon's patent for ``voice sniffing'' algorithms that analyze background audio for trigger words in order to supply users with targeted advertising and product recommendations~\cite{AmazonKeywordSniffer:uspat}. This functionality is in stark contrast to Amazon's current Terms of Use for Alexa, with audio transmission only occurring upon the detection of their device's wake word~\cite{AmazonAlexaFAQ}.

\subsection{Homegrown Bugs}
Privacy concerns over which data are actually being recorded and stored are not unfounded. In addition to the preeminent concern that stored user interactions (i.e. voice recordings and text transcripts) do not fall into the wrong hands, there are cases of extraneous recordings, recordings that don't fall under the aforementioned ``wake word'' behavior, being sent to external servers. For example, one journalist testing an early review unit of Google's Home Mini found the device recording and transmitting all audio to Google's servers around-the-clock. The issue was resolved by permanently disabling the device's touch-sensitive controls just before its official market release~\cite{AndriodPolice1}. Other cases have seen Amazon's Echo assistant mistaking a word for its wake word, followed by a  misinterpreted ``send message'' command, which subsequently sent private conversations to an address book contact~\cite{theguardian1}.

\subsection{Legislative Uncertainty}
Another area of major privacy concern lies in the unknown legislative capabilities for accessing user data. Data collected from Internet of Things (IoT) devices are more frequently surfacing as criminal evidence in court trials. In a 2017 murder case, data collected from e-mail time stamps, home security alarm logs, and a Fitbit tracker was used as evidence in charging the suspect~\cite{CNN1}. October 2018 saw the first case in the United States of a court ordering surveillance footage from Google's Nest smart home appliance subsidiary~\cite{FORBES}. A transparency report from Nest labs details approximately 300 user information requests from governments and courts between June 2015 to June 2018. The report states that search warrants will first be analyzed to rule out overly broad requests and ensure the requested data pertains only to the corresponding warrant before handing over data~\cite{NESTTR}.  The usage of connected device data in judicial trials has also recently extended to include smart speakers.

Shortly after Amazon's Alexa-enabled smart devices arrived on the U.S. market in 2015, an Arkansas court ordered Amazon to share recordings from a murder suspect's Echo speaker. While Amazon initially declined to provide the court with information, the suspect eventually provided consent to use the data as evidence~\cite{NPR}. More recently, Amazon has again been ordered to deliver Echo recordings for use as incriminating evidence~\cite{TechCrunch}. At the time of this writing, it remains unknown how Amazon will respond to the court's demands.

\subsection{Corporate Transparency}
Furthermore, it remains unknown how many general requests from courts and government bodies Amazon receives for Echo devices in general. This is due to their practice of listing requests across all brand divisions as a whole when publishing transparency reports~\cite{ZDnet1}. Google takes the same approach by not including per-device requests in its transparency reports, however it does include a ratio of the accounts received to the number of applicable accounts affected~\cite{ZDnet2}. In contrast, Apple does not publish any data request information for its HomePod smart speaker, stating there is no effective data to release because of its implementation of local differential privacy. Instead of associating server-destined user interactions with a user's account, differential privacy assign interactions to a random identifier~\cite{ApplePrivacy}. Research into Apple's implementation of differential privacy in macOS and iOS has shown a significant lack of transparency in the implementation among other shortcomings~\cite{tang2017privacy}.

An additional security vulnerability with the HomePod is its inability to differentiate between users, meaning anyone within the vicinity may request potentially sensitive information, such as personal notes and audio transcriptions of the primary user's text messages~\cite{voicebot_ai2}. Apple has filed a patent to implement user profiling but HomePod users, assuming they are aware of this potential for misuse, must currently choose between requiring manual authentication on their phone for every message dictation request, or accepting that anyone with access to the speaker: house guests, children, etc. can access their text messages until voice profiling capabilities have been implemented~\cite{AppleUserProfiling:uspat}.

How smart speaker manufacturers evaluate external data requests, how they store user data, how user Terms of Service are specified and updated, and what information is included in transparency reports are all entirely dependent upon the manufacturer and vary between each one. With software services becoming increasingly interconnected through the incorporation of third-party applications, e.g. Alexa Skills, it becomes exponentially difficult for users to maintain an overview of their personal data. As suggested by Lau et al., a legally binding set of industry standards (\`{a} la IEEE) and certifications that apply to all smart speaker and IoT device manufacturers is necessary to resolve the existing segmentation and ambiguity of company policy and legislature~\cite{lau2018alexa}.

\subsection{Speaker Vulnerability Exploitation}
Corporate transparency and legislative reach are not the only domains for which security must be considered. Hardware and software vulnerabilities must also be explored as points of exploitation. Research conducted in 2017 showed the Amazon Echo's UART connection port as a point of attack by enabling access to the device's firmware, thereby giving root shell access to its operating system. By installing a persistent shell script that launches when the device boots and exploiting Amazon's audio buffering application tool, raw microphone data could be streamed to the remote server of an attacker's choosing~\cite{barnes2017mwrlabs}. Although this method of attack required soldering an SD-card reader to the Echo, another research group proposed using a discrete 3D-printed attachment to mitigate evidence of tampering~\cite{clinton2016survey}. Amazon moved the +3V input pad to the main board for its second generation of Echo devices, meaning only 2015-2016 first generation devices are vulnerable to the ``wiretap'' attack.

Another form of attack applies to all generations of Alexa devices and does not require physical access to the device. This method of attack, known as ``skill squatting'', relies on Amazon's natural voice recognition algorithms making misinterpretations when transcribing speech-to-text~\cite{kumar2018skill}. For example, a user wishing to interact with a trusted banking skill, e.g. ``Capital One'', may ask Alexa to ``open Capital One''. Alexa may interpret the user to have said ``Capital Won'', triggering a malicious third-party skill of the same name. The malicious skill could then be used to obtain information about the user~\cite{zhang2018understanding}. Zhang et al. demonstrate a similar form of attack by means of a malicious skill with the same name as an authentic skill, but appended with a common phrase or word, for example, ``open Capital One, please'' launches the malicious skill ``Capital One Please'' instead of the authentic ``Capital One'' skill~\cite{zhang2018understanding}.

Kumar et al. noted that Alexa's speech transcription algorithms are non-deterministic by repeatedly presenting identical audio files over a reliable network connection and observing varying outcomes in the speech-to-text results~\cite{kumar2018skill}. Limiting speech samples to the Nationwide Speech Project (NSP) database, they identified 24 ``squattable'' words, i.e. semantic interpretation errors resulting from words that Alexa misinterprets both frequently and consistently~\cite{kumar2018skill}. The squattable words were used to identify over 30 susceptible skills available on the Alexa Skills Store but that number, limited by the study's proof-of-concept data set, is only an initial indicator for potential current and future vulnerable skills. Using skill squatting, the group was able to successfully demonstrate phishing attacks and suggests the possibility of using malicious skills to steal a user's log-in credentials. By applying gender and geographic predicates, they also exposed an extension of the skill squatting attack, ``spear skill squatting'', thus demonstrating the ability to target specific demographics. With a precedence for political manipulation using targeted advertisement and the growing smart speaker market, spear skill squatting has the potential to become a new channel of social influence~\cite{wired2016election}. Also noteworthy is the limited set of best practices currently used on the Alexa Skill Store when compared to other app store ecosystems, such as allowing multiple skills to have identical names.

Even though great progress in the development of voice processing and machine learning algorithms used in personal voice assistants has been made, as technology progresses, new possibilities of interacting with AI assistants will emerge, extending the types of data being exchanged and increasing the importance of data security. A recently submitted patent from Amazon details a feature that could determine some of the physical characteristics or emotional states of a user and react accordingly. For example, if a user is recognized as having a sore throat, Alexa might recommend a particular brand of cough medicine and order it. The patent mentions the detection of health conditions: having a cold, thyroid issues, sleepiness and emotional states including happiness, joy, anger, sorrow, sadness, fear, disgust, boredom, and stress~\cite{AmazonEmoCharacteristics:uspat}. With Amazon's acquisition of online pharmacy PillPack, Inc. in June 2018, there are inherit benefits to the convenience of supplying customers with  medicine and pharmaceuticals~\cite{businessinsider}. However, due to the considerable value of medical information, even when compared to other highly sensitive data such as credit card and social security information~\cite{healthcareasia}, there is equally an inherit risk in handling such valuable data securely, responsibly, and transparently.

The segmented set of rules and policies between device manufacturers, an incomplete set of best practices for third-party applications/skills, uncertain legislature, and the future potential for data harvesting all exhibit the current deficit surrounding data privacy and smart speakers.

\section{The Smart\textsuperscript{2} Speaker Blocker}
\subsection{Threat Model}
Before discussing the design of Smart\textsuperscript{2}, we first want to specify the threat model we are attempting to protect against. Our goal is to protect against sensitive audio data leakage, either to the device manufacturer or a malicious external third party. We assume that either the device, the manufacturer's server, or both may be compromised or that they, either intentionally or unintentionally, may not work as advertised.  As such, our goal is to physically prevent audio data we deem sensitive from ever reaching the microphone of the smart speaker and entering the unsafe environment.  However, once audio data does reach the smart speaker microphone, Smart\textsuperscript{2} provides no protection, nor do we provide any protection for any data stored on manufacturer servers. We also provide no protection against insider threats who have physical access to the Smart\textsuperscript{2} and may easily disable it.  Similarly the Smart\textsuperscript{2} filter is a tool to enforce a security policy for data leakage specified by the end user.  We provide no protection in the case that the end user does not specify his personal security policy correctly.  

\subsection{Design}
\begin{figure*}[ht]

\centering
\includegraphics[width=0.8\textwidth]{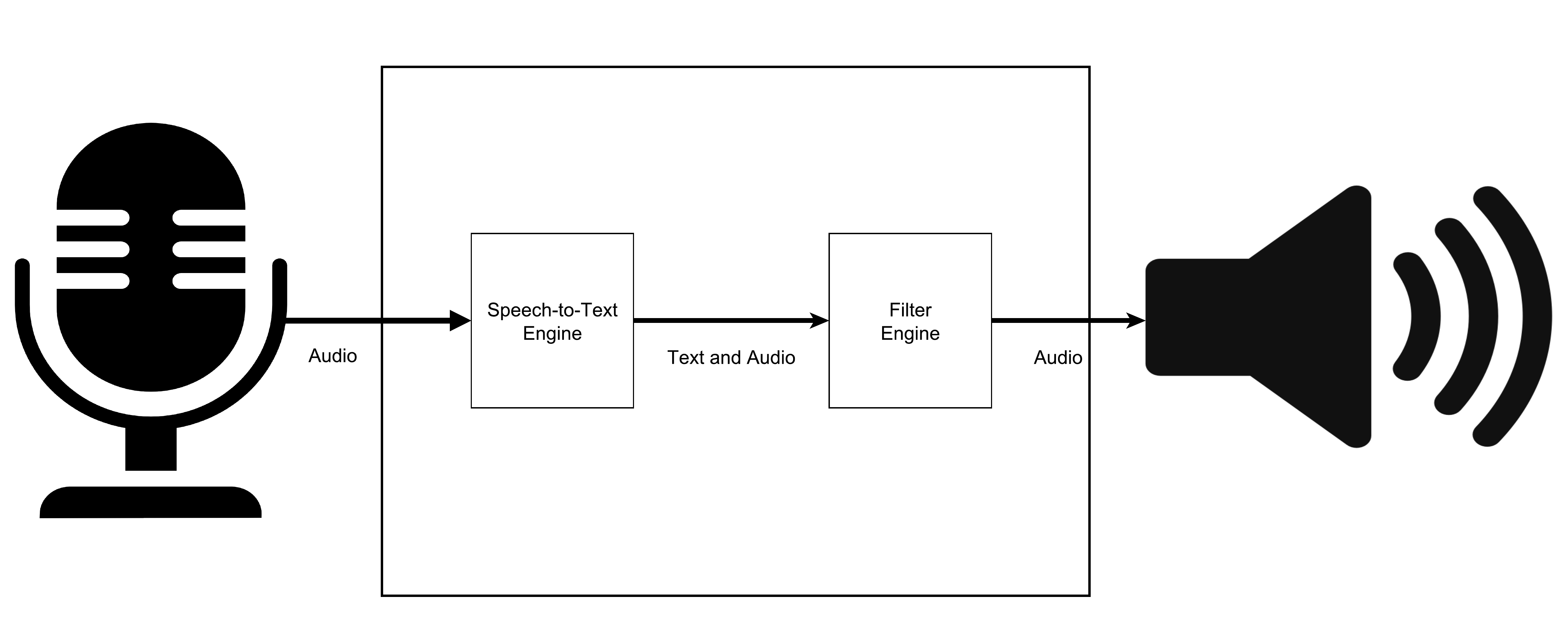}
\caption{Audio data comes in from the microphone and is converted to text by the speech-to-text engine.  Text and audio is both passed to the filtering engine, which decides whether or not to play the audio back for the smart speaker.}
\label{smartDiagram2}
\end{figure*}

The Smart\textsuperscript{2} is an intermediary that sits between the end user and the smart speaker, providing user-configured filtering to prevent sensitive audio from being recorded by the smart speaker.  It does this by enclosing the smart speaker microphone in soundproof material, along with an external speaker that is connected to the Smart\textsuperscript{2}.  The Smart\textsuperscript{2} has its own external microphone that it uses to record environmental audio.  This audio is fed into a speech-to-text engine in chunks, where each chunk is separated by a pause, indicating a phrase.  The speech-to-text engine converts this phrase to text, and passes it to the filtering engine.  The filtering engine uses a series of user-specified keywords and actions to decide whether to pass the phrase along to the smart speaker, drop the phrase, or take other action.  In order to handle different use cases, the Smart\textsuperscript{2} has several actions that it can take which are discussed in Section \ref{operation}.

The Smart$^2$ is based on open-source software and non-proprietary, off-the-shelf hardware. The main hardware components consist of a mini-ITX PC running Linux Ubuntu LTS 18.04, an external microphone for speech input, and a speaker for speech output to the smart speaker. A Python 3 virtual environment was then created to run Smart$^2$'s speech recognition software, facilitated by the SpeechRecognition Python library. This library was chosen for its ease of integration and support of both online and offline speech recognition engines. To improve security the Smart$^2$ is an air-gapped device, so we used an offline speech recognition engine developed by researchers at Carnegie Mellon University to ensure that interactions with the device remain local. CMU's CMUSphinx, an open source, Hidden Markov Model (HMM) based speech recognition system, was determined to be the most viable solution due to its offline nature~\cite{Walker04sphinx-4:a}. For our system, we used the CMUSphinx US English acoustic model for continuous speech, the US English generic language model, and CMUDict, a large vocabulary dictionary.

\subsection{Limitations of the Prototype}
Most smart speakers include multiple microphones so that they can hear incoming audio from any direction, regardless of how they are placed.  Because this is a prototype system, we did not implement this feature, and instead used a single omnidirectional microphone and placed it against a wall.  Also, while we recognize that hardware cost and power consumption are issues for a production system, we focus more on the system design and usability, and note that while our hardware is more expensive and consumes more power than a production system would, the software itself used only a small amount of memory and processor power and could easily be adapted to an embedded system.  Additionally, for extremely low spec systems, there is an embedded version of the CMUSphinx system, called PocketSphinx, that was designed to run on systems as low end as a 200Mhz ARM processor with 16MB of memory \cite{huggins2006pocketsphinx}.

\subsection{Offline vs Online Speech Recognition}
One concern with an offline system like the Smart$^2$ Speaker Blocker is the accuracy of the offline speech recognition engine as compared to an online engine, like those used by smart speakers themselves.  While the CMUSpinx software itself is not as accurate as, for example, the online Google speech recognition software, Google has recently developed an offline speech-to-text system that is as accurate as its online system, only requires 80MB of memory, can run on low-end mobile devices, and operates faster-than-realtime on a single core \cite{GoogleOfflineSpeech}.  While this system is currently only available as part of their GBoard keyboard software and has not been released for public use, it demonstrates that it is feasible for a low-end device such as the Smart$^2$ Speaker Blocker to recognize speech as quickly and accurately as an online system.   
\subsection{Operation}
\label{operation}

The Smart\textsuperscript{2} can be configured by the user by using pattern/action pairs.  A pattern can simply be a keyword, such as ``password'', or for more sophisticated users it can include regular expressions. In addition, some actions can take additional parameters.  The current set of actions includes:

\begin{itemize}
    \item \textbf{Phrase filter} - The phrase filter action drops the current phrase.  This action is intended to be used in cases where the current phrase may have some sensitive information, but we don't expect the information to extend past the current phrase.  For example, we may use the pattern ``password'' with the phrase filter action, because we want to filter cases where the user says things like ``My password is XXX'', but we don't believe more sensitive information will be discussed after this phrase.
    \item \textbf{Timed filter} - The timed filter takes an additional duration parameter, and does not transmit audio for a fixed duration after detecting the pattern.  This is intended to filter more sensitive discussions that may extend beyond the current phrase.  For example, the pattern may be set to ``girlfriend'' and the duration to 10 minutes, because we do not want sensitive personal discussions transmitted to the smart speaker, and we expect that this will be part of a discussion.  The duration of this filter may be extended if other timed filter patterns are detected, and the duration is the time to wait after the last occurrence of the pattern.
    \item \textbf{Pattern deletion filter} - The pattern deletion filter deletes the pattern from the phrase, and transmits it to the smart speaker.  It does this by removing the pattern from the phrase text, and using a text-to-speech engine to play back the modified text.
\end{itemize}

In addition to pattern matching, the Smart\textsuperscript{2} has several other modes that are activated by keywords or other triggers.  These are:

\begin{itemize}
    \item \textbf{Privacy mode} - This mode is activated by keyword, and does not transmit any data to the smart speaker until it is turned off.  While this mode replicates the physical ``Mute'' button on most smart speakers, for many smart speakers it is not clear if the mute button is a hardware mute button, meaning no electrical signals are sent from the microphone to the device, or a software mute, meaning the software ignores the input from the microphone.  In many cases it is strongly suspected that the button is a software button, in which case the software may intentionally or through programming errors continue to transmit.  Software mute buttons are also vulnerable to exploits which can render them inoperable.  Privacy mode guarantees (to the extent that Smart\textsuperscript{2} is bug-free) that sound cannot be recorded by the smart speaker.
    
    \item \textbf{Strict hotword mode} - In this mode \textit{only} audio that is prefixed by the hotword for the smart speaker is passed through.  All other audio is blocked.  This completely prevents the smart speaker from inadvertently listening to any conversation that is not a specific command.
    
    \item \textbf{Loudness mode} - Loudness mode uses a decibel estimator to estimate the loudness of the speaker's voice.  This mode is an example of a more general emotion detection filter, where the user can filter based on their current mood.  Loudness mode is designed to detect arguments or disagreements, which the user is unlikely to want transmitted to the smart speaker. This mode was inspired by a recent Amazon patent which detects the mood of smart speaker users \cite{AmazonEmoCharacteristics:uspat}.
    
    \item \textbf{Text-to-speech mode} - This mode is activated by keyword, and uses text-to-speech to mask the identity, gender, or other identifying features of the speaker, providing an extra layer of anonymity.  
\end{itemize}

\subsection{Configuration}
The Smart\textsuperscript{2} speaker filter is completely configurable by the user. In the current prototype the set of filters to be used, keywords, wait times, and other parameters for the filters are stored in a plain text format.  In a production system this would be replaced by a simple graphical interface for ease-of-use.  To aid novice users, a default policy configuration consisting of common sensitive phrases could be included that the user could customize.

\section{Performance Evaluation}
In our analysis of the Smart\textsuperscript{2} system we consider three aspects of the system.  The first aspect is system performance impact, meaning the impact of the Smart\textsuperscript{2} system on the performance of the smart speaker from a latency perspective. The second part attempts to address the increased security offered by the Smart\textsuperscript{2}.  To measure this, we measure the amount of sensitive information that is blocked by the Smart$^2$ Speaker Blocker.  The third part of our evaluation is to consider the usability of the system, both with respect to the ease of configuration and the impact on the usability of the smart speaker itself.

\subsection{System Performance Impact}
To evaluate the impact of the Smart\textsuperscript{2} on smart speaker performance, we consider the latency introduced to the smart speaker responses due to the Smart\textsuperscript{2} and the increases in the smart speaker error rate both in general and when used with the additional speech-to-text layer. 
\begin{figure*}[ht]

\centering
\includegraphics[width=0.8\textwidth]{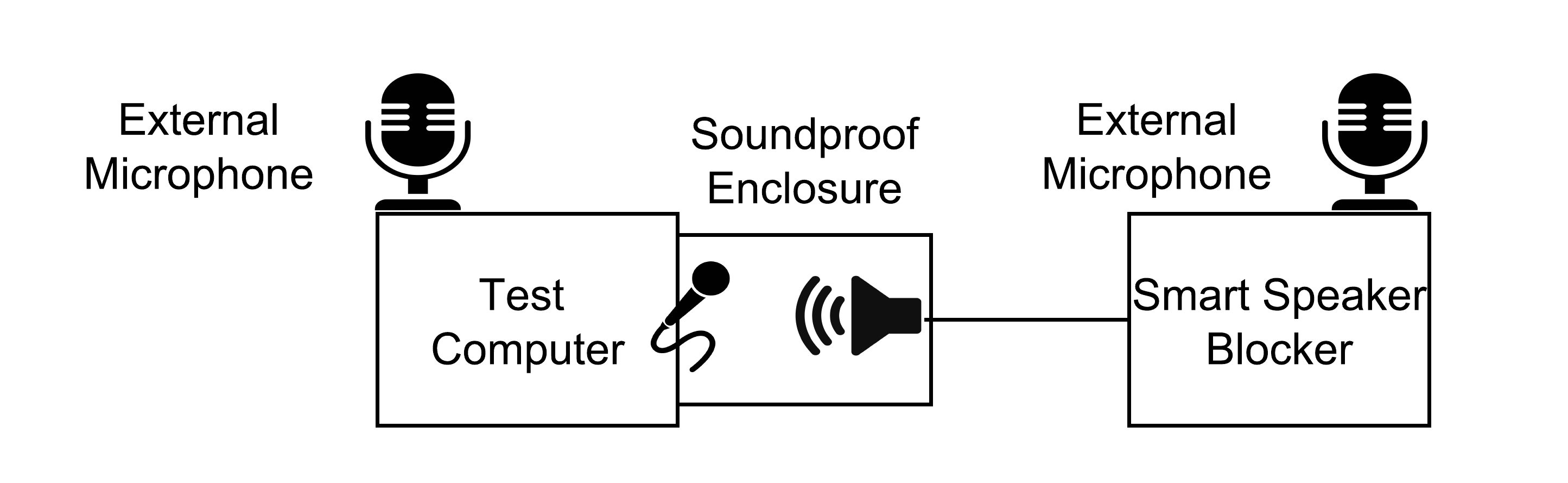}
\caption{Our test setup consists of the Smart\textsuperscript{2} along with a separate networked computer which records both the raw audio and the output from the Smart\textsuperscript{2}.  Both sets of audio recordings were transcribed using the online Google speech-to-text API, and compared.}
\label{smartDiagram3}
\end{figure*}
\subsubsection{Experimental Setup}
The hardware used for our tests consists of a mini-ITX PC with an i7 7500u 15W mobile CPU, 8GB of memory, and a 128GB SSD drive running Linux Ubuntu LTS 18.04, an external microphone for speech input, and a speaker for speech output to the smart speaker.  The system was developed in Python 3, using the SpeechRecognition package and the CMUSphinx library with the CMUDict dictionary for US English for speech recognition.  Text-to-speech was done using the PyTTSX3 library, an offline library. The smart speaker used was the Amazon Echo Dot, Generation 2.  

\subsubsection{Latency}
The Smart\textsuperscript{2} system by necessity introduces some latency into smart speaker interactions, simply because it must play back audio. This is somewhat mitigated by the fact that most smart speaker commands are generally quite short, for example querying about the weather or time. In addition to the playback time, additional latency is introduced by the speech-to-text (STT) engine and for some modes, the text-to-speech (TTS) engine.  Because the playback time is fixed, and always equal to the length of the phrase spoken, we focus our measurements on the processing overhead introduced.  Additionally, while there are many filters available, they all utilize the speech-to-text engine in the same way, so we present one set of results rather than one set per filter.  The one exception to this is the filters which utilize text-to-speech, however in our tests we found that the computation time required for text-to-speech was negligible.

We evaluated Smart$^2$'s processing overhead by running a controlled experiment using four separate speech recordings as input data for the system. Each test was timed using the system's current time at the end of the capture audio procedure. Each test was repeated ten times and the results averaged. The tests were conducted by the system as follows:
\begin{enumerate}
\itemsep0em
    \item Initialize microphone for audio input
    \item Capture ambient background noise and adjust the microphone accordingly (SpeechRecognition)
    \item Play recording
    \item Capture audio data
    \item Retrieve system time
    \item Transcribe speech to text (CMUSphinx)
    \item Evaluate text transcript via the filter engine
    \item Write elapsed time, evaluation results, and transcription to file output
\end{enumerate}

\begin{table}[!t]
	\centering
	\renewcommand{\arraystretch}{1.0}

	\caption{ Processing Times (ms)}
	\label{table_example}
	\begin{tabular}{c|c|c|c|c}
		\hline
		{\scriptsize \textbf{Input}} & {\scriptsize \textbf{Action}} & {\scriptsize \textbf{Min} } & {\scriptsize \textbf{Max}} & {\scriptsize \textbf{Avg (\textit{/10})}}\\
		\hline
	   {\scriptsize	``How's the weather today?''} & {\scriptsize FWD } & {\scriptsize 205} & {\scriptsize 227 } & {\scriptsize 210}\\
		\hline
		{\scriptsize ``Bank account password.''} & {\scriptsize BLOCK } & {\scriptsize 197 }  & {\scriptsize 228 } & {\scriptsize 205}\\
		\hline
		{\scriptsize ``What's 10 x 3?''} & {\scriptsize FWD } & {\scriptsize 186 } & {\scriptsize 201 } & {\scriptsize 192}\\
		\hline
		{\scriptsize ``Play Pachelbel Canon in D.''} & {\scriptsize FWD } & {\scriptsize 218 } & {\scriptsize 240 } & {\scriptsize 226}\\
		\hline
	\end{tabular}
\end{table}
The process was repeated ten times for each recording. Performance testing results are documented in Table I, showing the minimum, maximum, and average processing time in milliseconds.  Testing the system indicated the Smart\textsuperscript{2} adds approximately 200 milliseconds of overhead to convert speech to text and apply the filtering rules.  The vast majority of this time is used by the speech-to-text library, with the filtering process taking an almost undetectable amount of time due to the efficiency of the Python regex library and the small text size.  We additionally repeated these tests including the time to convert the output text to audio using the text-to-speech library, but found that the additional overhead introduced by this step was negligible.

\subsubsection{Accuracy}
The error rate of the speech-to-text library is determined by the size of the dictionary, the complexity of the model, and various other parameters.  The Sphinx4 library has reported error rates as low as 3.9\%, for small dictionary sizes\cite{Walker04sphinx-4:a}.  In comparison, server-based speech-to-text services, like those used by smart speakers, typically achieve around 5\% error rates for general speech\cite{DBLP:journals/corr/abs-1708-06073}. 

To test the accuracy of the Spinx4 library used in our system, we recorded a series of phrases and conversations.  We then chose three environments to test: 
\begin{itemize}
    \item An empty, quiet room
    \item A room with a television on in the background
    \item A room with a recording of four people holding a conversation playing in the background
\end{itemize}
We played back our pre-recorded phrases and conversations in each environment and transcribed them using both Google online text-to-speech (default model) and the Sphinx4 library, then compared the transcripts with a hand transcription. We repeated this ten times and averaged the results. The error rates for the Google online engine in our test were 4.1\%, 4.1\%, and 6.4\%, respectively.  For Sphinx4 the error rates were 5.6\%, 7.2\%, and 7.5\%.  

While the Sphinx4 engine had higher error rates than the Google online engine, the results were reasonable. Anecdotally, errors were typically in more complex words, proper nouns, and homophones.  The Sphinx4 engine is much older than the Google engine, so these results were expected.  However, as we mentioned in section 3.4, offline engines can perform as well as online engines.

\subsection{Experimental Review and Participant Selection}
Because our experiments in the next section involved human subjects, our experimental design was reviewed and approved by our University Research Ethics Subcommittee (our equivalent of the Institutional Review Board (IRB)).  Some concerns were raised regarding our storage of personal data, particularly recordings or transcripts of conversations.  Because of this, we verified that no audio recordings or transcripts were stored in any of our experiments. We store only counts of policy violations, rather than transcripts of the conversations.  Because the configuration files themselves may contain sensitive information, including words the user considered sensitive, these files were deleted by the participants prior to returning the devices.

Participants in our experiments were volunteers recruited from the University via an e-mail advertisement.  All had prior experience using smart speakers.  Participants were given an introduction to how the Smart$^2$ functioned and told that it would be constantly listening to their conversations, but would not be recording or transcribing them.  They were also trained in how to configure the various filter options and given a manual and example configuration.  As compensation for participation, volunteers were given a new Amazon Echo Dot.

\subsection{Assessment}
\label{security}
To assess security and usability we setup the Smart\textsuperscript{2} in a test configuration in ten home environments with a total of 25 users where the home owners regularly use a smart speaker.  We assisted the home owners in setting up and configuring the Smart$^2$ for use with an Amazon Echo Dot that we also supplied. After training, users configured the Smart$^2$ to meet their personal privacy preferences. The Smart$^2$ Speaker was used in each home for two weeks. At the end of the test, the device was collected and all family members were given a survey and interviewed regarding their experiences with the device. 

Our test configuration, shown in Figure \ref{smartDiagram3}, consisted of the Smart\textsuperscript{2} and a separate computer with two microphones.  One microphone was used to record output from the Smart\textsuperscript{2} and the other to record the raw, unfiltered audio.  Both the raw audio and filtered audio were recorded and converted to text using the online Google speech-to-text API, which we used as our ground truth.  The software then compared the text with the sensitive word list from the Smart$^2$ device configuration to determine how many instances of privacy sensitive words were in the original audio vs the filtered audio.   For testing purposes, both computers were connected to each other via a private wired network, allowing the test computer to access the Smart\textsuperscript{2} filter configuration file, which was used to calculate statistics.  As per our IRB recommendations, we only recorded that sensitive words were filtered or not, and not the word being filtered. For failure cases, we additionally recorded the number and type of the rule that should have triggered the word to be filtered. No copies of transcripts or recordings were ever written to disk, and the configuration files were deleted from both machines at the end of the testing period.  Only aggregate statistics were kept.

\begin{table*}[t]
 	\centering

	\caption{This table shows a breakdown of the various filter usage. Minimum and Maximum are the minimum and maximum number of that filter type that was configured by a user.  Total Activations is total number of times across all users that filter was triggered to filter voice data.  Total Failures is the total number of times across all users that filter should have been activated, but was not.}
 	\label{table_popularity}
 	\begin{tabular}{c|c|c|c|c|c|c}
 		\hline
 		
		{\ \textbf{ Filter Type }} & { \textbf{Average per User}} & { \textbf{Minimum}} & { \textbf{Maximum}} & { \textbf{Number of Users}} & { \textbf{Total Activations}} & {\textbf{Total Failures}} \\
		\hline
  	   {	Phrase filter} &            { 12 } & { 3 } & { 28 } & { 25} & { 689} & {65}\\
 		\hline
 		{ Timed filter} &            { 7 } & { 1 } & { 10 } & { 18} & { 233} & {21}\\
 		\hline
 		{ Pattern deletion filter} & { 4 } & { 3 } & { 15 } & { 15} & { 37} & {6}\\
 		\hline
 		{	Privacy mode} &         { - } & { - } & { - } & { 18} & { 97} & {0}\\
 		\hline
 		{	Strict hotword mode} &  { - } & { - } & { - } &  { 17} & { 0} & {0} \\
        \hline
 		{	Loudness mode} &        { - } & { - } & { - } & { 24} & { 102} & {0}\\
 		\hline
		{ Text-to-speech mode} &  {  - } & {  - } & {  - } & {  5 } & {  21 } & {0} \\
        \hline
 	\end{tabular}
\end{table*}

Statistics for each filter type are shown in Table \ref{table_popularity}.  For the three filters that can be defined for multiple sensitive words, we include average number defined per user who chose to use that filter type, disregarding users who did not create any of that kind of filter, as well as the minimum and maximum number of filters that any user chose to create. The other four filters are modal, and can either be turned on or off.  For all the filters we show the total number (out of 25) of users who chose define one or more of those filters, or to turn on that mode, the total number of activations (the number of times that filter was triggered), and the total number of failures, where the filter or mode should have been triggered but was not. 

The most popular filter was the phrase filter, followed by timed filters, and pattern deletion filters.  These filters were also the sources of all the activation failures.  While for privacy reasons we were unable to record the keyword that failed, we did record the ordinal number of the rule that failed.  From this data, we found that 85\% of the failures came from just 12\% of the rules, most likely from limitations of the recognition engine in recognizing certain words such as proper names.  This suggests that future implementations should incorporate some type of enhanced training mechanism specifically for the keywords, rather than the more general dictionary that we used.

Of the four modes, the loudness mode was the most popular by far with 24 out of 25 participants enabling it.  This was an unexpected result, but during the post-experiment interview several participants expressed that they felt this was one of the most useful features. The rational given for this was that they felt that inter-personal conflict between family members was one of the most private interactions in the home, and they were very uncomfortable with the idea of strangers overhearing them.  None of the modes failed to activate, presumably because the activation keywords selected were simple and easily detected by the speech-to-text engine.

Over the two week period, the test computer detected a total of slightly more than 1.2 million words, with an average phrase length of 12 words. The total number of sensitive phrases detected was 1121, with 959 of these being filtered and 92 failures, among the three filters.  The means that the Smart$^2$ filtered 91.8\% of sensitive data, a significant improvement over the unprotected smart speaker. In addition, the 4 enhanced privacy modes were engaged a total of 220 times, with no failures, further improving the privacy offered by the Smart$^2$.

To assess the usability of the device, we gave all 25 users a short six question Likert scale-based survey at the end of the experiment.  The questions included:
\begin{enumerate}
\item Did your smart speaker seem noticeably slower than usual? (1 - not at all to 5 - much slower)
\item Does the Smart$^2$ make you more comfortable talking near your smart speaker? (1 - not at all to 5 - much more comfortable)
\item Do you feel the Smart$^2$ made your smart speaker unusable? (1 - completely unusable to 5 - about the same)
\item Was the Smart$^2$ easy to configure? (1 - very difficult to 5 - very easy)
\item Do you feel the Smart$^2$ makes you safer? (1 - not safer at all to 5 - much safer)
\item What is your opinion of the Smart$^2$? (1 - very unfavorable to 5 very favorable)
\end{enumerate}

As can be seen in figure \ref{LikertPlot} most users noticed a small decrease in the speed of the smart speaker when used with the Smart$^2$, but did not feel it made the smart speaker unusable.  Users also overwhelmingly felt that the Smart$^2$ was easy to configure and improved the security of their smart speaker.  Correspondingly, the overall opinion of the Smart$^2$ was very favorable.   
\begin{figure}[h]
\centering
\includegraphics[width=0.45\textwidth]{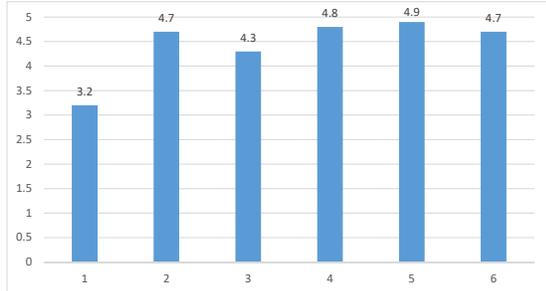}
\caption{Average ratings for Likert-based post-experiment survey.}
\label{LikertPlot}
\end{figure}
\section{Discussion and Future Work}
Smart speakers pose an enormous potential privacy risk to consumers, a risk that they currently have little way to mitigate.  In this paper we present the Smart\textsuperscript{2} system, which leverages the same machine learning algorithms that endanger user privacy to offer transparent, open-source protection for users of smart speakers that is completely under their control.  We then demonstrate that the device offers significant privacy enhancement, filtering more than 91.8\% of sensitive audio in our test, while at the same time minimally impacting the performance of the smart speaker.  

While the Smart\textsuperscript{2} is a complete system as presented, we feel there is substantial work remaining to be done.  The most substantial improvement would be to replace the older Sphinx4 engine with the newly developed Google offline speech-to-text engine, which would increase accuracy and decrease latency to near zero. Other improvements would be implementing emotion detection based filtering and other filtering modes that are not strictly keyword based.  We also feel that approach of using machine learning to develop systems that protect or enhance the security of end user systems and devices is a promising technique that has not been broadly explored either commercially or in the research literature.





\bibliographystyle{IEEEtranS}

\bibliography{references}

\begin{thebibliography}{10}
\providecommand{\url}[1]{#1}
\csname url@samestyle\endcsname
\providecommand{\newblock}{\relax}
\providecommand{\bibinfo}[2]{#2}
\providecommand{\BIBentrySTDinterwordspacing}{\spaceskip=0pt\relax}
\providecommand{\BIBentryALTinterwordstretchfactor}{4}
\providecommand{\BIBentryALTinterwordspacing}{\spaceskip=\fontdimen2\font plus
\BIBentryALTinterwordstretchfactor\fontdimen3\font minus
  \fontdimen4\font\relax}
\providecommand{\BIBforeignlanguage}[2]{{%
\expandafter\ifx\csname l@#1\endcsname\relax
\typeout{** WARNING: IEEEtranS.bst: No hyphenation pattern has been}%
\typeout{** loaded for the language `#1'. Using the pattern for}%
\typeout{** the default language instead.}%
\else
\language=\csname l@#1\endcsname
\fi
#2}}
\providecommand{\BIBdecl}{\relax}
\BIBdecl

\bibitem{AndriodPolice1}
AndroidPolice, ``Google is permanently nerfing all {H}ome {M}inis because mine
  spied on everything {I} said 24/7,'' 2018,
  \url{https://www.androidpolice.com/2017/10/10/google-nerfing-home-minis-mine-spied-everything-said-247/#2}
  [Online; accessed December 3, 2018].

\bibitem{ApplePrivacy}
Apple, ``Approach to privacy,'' 2018,
  \url{https://www.apple.com/privacy/approach-to-privacy/} [Online; accessed
  January 3, 2019].

\bibitem{BigbroAwards1}
B.~B. Awards, ``Watching the watchmen worldwide,'' 2018,
  \url{http://www.bigbrotherawards.org/} [Online; accessed November 7, 2018].

\bibitem{barnes2017mwrlabs}
\BIBentryALTinterwordspacing
M.~Barnes. (2017) {A}lexa, are you listening?
  \url{https://labs.mwrinfosecurity.com/blog/alexa-are-you-listening} [Online;
  accessed October 15, 2018]. [Online]. Available:
  \url{https://labs.mwrinfosecurity.com/blog/alexa-are-you-listening}
\BIBentrySTDinterwordspacing

\bibitem{FORBES}
T.~Brewster, ``Smart home surveillance: Governments tell {G}oogle's {N}est to
  hand over data 300 times,'' 2018,
  \url{https://www.forbes.com/sites/thomasbrewster/2018/10/13/smart-home-surveillance-governments-tell-googles-nest-to-hand-over-data-300-times/#2b7aa83e2cfa}
  [Online; accessed December 18, 2018].

\bibitem{canalys}
Canalys, ``Smart speaker installed base to hit 100 million by end of 2018,''
  2018,
  \url{https://www.canalys.com/newsroom/smart-speaker-installed-base-to-hit-100-million-by-end-of-2018}
  [Online; accessed November 11, 2018].

\bibitem{businessinsider}
J.~Ciolli, ``Amazon’s \$1 billion purchase of {P}ill{P}ack wiped out 15 times
  that from pharmacy stocks — and it shows the outsize effect the juggernaut
  can have on an industry,'' 2018,
  \url{https://www.businessinsider.sg/amazon-pharmacy-pillpack-acquisition-merger-showing-outsized-impact-2018-6/?r=US&IR=T}
  [Online; accessed November 25, 2018].

\bibitem{clinton2016survey}
I.~Clinton, L.~Cook, and S.~Banik, ``A survey of various methods for analyzing
  the {A}mazon {E}cho,'' 2016,
  \url{https://vanderpot.com/Clinton_Cook_Paper.pdf} [Online; accessed October
  16, 2018].

\bibitem{CNN1}
CNN, ``Cops use murdered woman's {F}itbit to charge her husband,'' 2018,
  \url{https://edition.cnn.com/2017/04/25/us/fitbit-womans-death-investigation-trnd/index.html}
  [Online; accessed December 18, 2018].

\bibitem{BigbroAwards2}
B.~B.~A. DE, ``Consumer protection: {A}mazon {A}lexa,'' 2018,
  \url{https://bigbrotherawards.de/kategorie/verbraucherschutz} [Online;
  accessed November 7, 2018].

\bibitem{NPR}
C.~Dwyer, ``Arkansas prosecutors drop murder case that hinged on evidence from
  {A}mazon {E}cho,'' 2018,
  \url{https://www.npr.org/sections/thetwo-way/2017/11/29/567305812/arkansas-prosecutors-drop-murder-case-that-hinged-on-evidence-from-amazon-echo}
  [Online; accessed November 30, 2018].

\bibitem{AmazonKeywordSniffer:uspat}
K.~K. Edara, ``Key word determinations from voice data,'' U.S. Patent
  8\,798\,995, Aug. 5, 2014.

\bibitem{theguardian1}
T.~Guardian, ``{A}mazon's {A}lexa recorded private conversation and sent it to
  random contact,'' 2018,
  \url{https://www.theguardian.com/technology/2018/may/24/amazon-alexa-recorded-conversation}
  [Online; accessed December 3, 2018].

\bibitem{AppleUserProfiling:uspat}
A.~P. Haughay, ``User profiling for voice input processing,'' U.S. Patent
  10\,049\,675, Aug. 14, 2018.

\bibitem{healthcareasia}
\emph{Health data is wealth: Why hackers targeted {S}ingapore}, vol.
  September-February 2019, pp. 12-13, Health Care Asia Magazine, Sep. 2018,
  \url{https://healthcareasiamagazine.com/sites/default/files/healthcareasiamagazine/print/HCA\_BODY-\%2012-13.pdf}
  [Online; accessed November 15, 2018].

\bibitem{huggins2006pocketsphinx}
D.~Huggins-Daines, M.~Kumar, A.~Chan, A.~W. Black, M.~Ravishankar, and A.~I.
  Rudnicky, ``Pocketsphinx: A free, real-time continuous speech recognition
  system for hand-held devices,'' in \emph{Acoustics, Speech and Signal
  Processing, 2006. ICASSP 2006 Proceedings. 2006 IEEE International Conference
  on}, vol.~1.\hskip 1em plus 0.5em minus 0.4em\relax IEEE, 2006, pp. I--I.

\bibitem{AmazonEmoCharacteristics:uspat}
H.~Jin and S.~Wang, ``Voice-based determination of physical and emotional
  characteristics of users,'' U.S. Patent 10\,096\,319, Oct. 09, 2018.

\bibitem{voicebot_ai2}
B.~Kinsella, ``{A}pple {H}ome{P}od has a privacy flaw that no one is talking
  about,'' 2018,
  \url{https://voicebot.ai/2018/02/11/apple-homepod-privacy-flaw-no-one-talking/}
  [Online; accessed December 12, 2018].

\bibitem{voicebot_ai1}
------, ``{U.S.} smart speaker users rise to 57 million,'' 2018,
  \url{https://voicebot.ai/2018/10/30/u-s-smart-speaker-users-rise-to-57-million/}
  [Online; accessed December 13, 2018].

\bibitem{kumar2018skill}
D.~Kumar, R.~Paccagnella, P.~Murley, E.~Hennenfent, J.~Mason, A.~Bates, and
  M.~Bailey, ``Skill squatting attacks on {A}mazon {A}lexa,'' in \emph{27th
  $\{$USENIX$\}$ Security Symposium ($\{$USENIX$\}$ Security 18)}, 2018, pp.
  33--47.

\bibitem{wired2016election}
I.~Lapowsky, ``How {R}ussian {F}acebook ads divided and targeted {US} voters
  before the 2016 election,'' 2018,
  \url{https://www.wired.com/story/russian-facebook-ads-targeted-us-voters-before-2016-election/}
  [Online; accessed November 23, 2018].

\bibitem{lau2018alexa}
J.~Lau, B.~Zimmerman, and F.~Schaub, ``{A}lexa, are you listening?: Privacy
  perceptions, concerns and privacy-seeking behaviors with smart speakers,''
  \emph{Proceedings of the ACM on Human-Computer Interaction}, vol.~2, no.
  CSCW, p. 102, 2018.

\bibitem{NESTTR}
Nest, ``Transparency report: Requests for user information,'' 2018,
  \url{https://nest.com/legal/transparency-report/} [Online; accessed December
  19, 2018].

\bibitem{NPR_Edison}
NPR and E.~Research, ``The smart audio report, spring 2018,'' 2018,
  \url{https://www.nationalpublicmedia.com/wp-content/uploads/2018/07/Smart-Audio-Report-from-NPR-and-Edison-Research-Spring-2018_Downloadable-PDF.pdf}
  [Online; accessed November 15, 2018].

\bibitem{GoogleOfflineSpeech}
J.~Schalkwyk, ``An all-neural on-device speech recognizer,'' 2019,
  \url{https://ai.googleblog.com/2019/03/an-all-neural-on-device-speech.html}
  [Online; accessed May 13, 2019].

\bibitem{AmazonAlexaFAQ}
A.~C. Service, ``{A}lexa and {A}lexa device {FAQ}s,'' 2018,
  \url{https://www.amazon.com/gp/help/customer/display.html?nodeId=201602230}
  [Online; accessed November 9, 2018].

\bibitem{tang2017privacy}
J.~Tang, A.~Korolova, X.~Bai, X.~Wang, and X.~Wang, ``Privacy loss in {A}pple's
  implementation of differential privacy on mac{OS} 10.12,'' \emph{arXiv
  preprint arXiv:1709.02753}, 2017.

\bibitem{Walker04sphinx-4:a}
W.~Walker, P.~Lamere, P.~Kwok, B.~Raj, R.~Singh, E.~Gouvea, P.~Wolf, and
  J.~Woelfel, ``Sphinx-4: A flexible open source framework for speech
  recognition,'' Tech. Rep., 2004.

\bibitem{ZDnet1}
Z.~Whittaker, ``{A}mazon turns over record amount of customer data to {US} law
  enforcement,'' 2018,
  \url{https://www.zdnet.com/article/amazon-turns-over-record-amount-of-customer-data-to-us-law-enforcement/}
  [Online; accessed December 9, 2018].

\bibitem{ZDnet2}
------, ``{A}mazon won't say if it hands your {E}cho data to the government,''
  2018,
  \url{https://www.zdnet.com/article/amazon-the-least-transparent-tech-company/}
  [Online; accessed January 3, 2019].

\bibitem{TechCrunch}
------, ``Judge orders {A}mazon to turn over {E}cho recordings in double murder
  case,'' 2018,
  \url{https://techcrunch.com/2018/11/14/amazon-echo-recordings-judge-murder-case/}
  [Online; accessed December 1, 2018].

\bibitem{DBLP:journals/corr/abs-1708-06073}
\BIBentryALTinterwordspacing
W.~Xiong, L.~Wu, F.~Alleva, J.~Droppo, X.~Huang, and A.~Stolcke, ``The
  {M}icrosoft 2017 conversational speech recognition system,'' \emph{CoRR},
  vol. abs/1708.06073, 2017. [Online]. Available:
  \url{http://arxiv.org/abs/1708.06073}
\BIBentrySTDinterwordspacing

\bibitem{zhang2018understanding}
N.~Zhang, X.~Mi, X.~Feng, X.~Wang, Y.~Tian, and F.~Qian, ``Understanding and
  mitigating the security risks of voice-controlled third-party skills on
  {A}mazon {A}lexa and {G}oogle {H}ome,'' \emph{arXiv preprint arXiv
  :1805.01525}, 2018.

\end{thebibliography}

\end{document}